\documentclass[twocolumn]{aastex61}
\usepackage{url}
\usepackage{epstopdf}
\usepackage{enumitem} 
\usepackage{amsmath}

\newcommand{\hMsun}{{\ifmmode{h^{-1}{\rm
        {M_{\odot}}}}\else{$h^{-1}{\rm{M_{\odot}}}$~}\fi}} 
\newcommand{\hMpc}{{\ifmmode{h^{-1}{\rm Mpc}}\else{$h^{-1}$Mpc }\fi}}
\def\be{\begin{equation}}
\def\ee{\end{equation}}
\def\ba{\begin{eqnarray}}
\def\ea{\end{eqnarray}}

\received{}
\revised{}
\accepted{}
\submitjournal{ApJS}

\begin{document}
\title[Graph Database Spatial Statistics]{Graph Database Solution for Higher Order Spatial Statistics in the Era of Big Data}


\correspondingauthor{Juhan Kim}
\email{kjhan@kias.re.kr}

\author[0000-0002-5513-5303]{Cristiano G. Sabiu}
\affiliation{Department  of  Astronomy,  Yonsei  University,  50  Yonsei-ro, Seoul 03722, Korea}

\author[0000-0002-2571-1357]{Ben Hoyle}
\affiliation{Universitaets-Sternwarte, Fakultaet fur Physik, Ludwig-Maximilians Universitaet Muenchen, Scheinerstr. 1, D-81679 Muenchen, Germany}
\affil{Max Planck Institute fur Extraterrestrial Physics, Giessenbachstr. 1, D-85748 Garching, Germany}

\author[0000-0002-4391-2275]{Juhan Kim}
\affil{Center for Advanced Computation, Korea Institute for Advanced Study, 85 Hoegi-ro, Dongdaemun-gu, Seoul 02455, Korea}

\author[0000-0003-3964-0438]{Xiao-Dong~Li}
\affil{School of Physics and Astronomy, Sun Yat-Sen University, Guangzhou 510297,  People's Republic of China}

\begin{abstract}
We present an algorithm for the fast computation of the general $N$-point spatial correlation functions of any discrete point set embedded within an Euclidean space of $\mathbb{R}^n$. Utilizing the concepts of kd-trees and graph databases, we describe how to count all possible $N$-tuples in binned configurations within a given length scale, e.g. all pairs of points or all triplets of points with side lengths $<r_{max}$.
Through bench-marking  we show the computational advantage of our new graph based algorithm over more traditional methods. We show that all 3-point configurations up to and beyond the Baryon Acoustic Oscillation scale ($\sim$200 Mpc in physical units) can be performed on current SDSS data in reasonable time. Finally we present the first measurements of the 4-point correlation function of $\sim$0.5 million SDSS galaxies over the redshift range $0.43<z<0.7$. 

\end{abstract}


\section{Introduction}
Cosmology is now entering an era of {\em big data}. From large optical surveys like LSST\footnote{https://www.lsst.org}, Euclid\footnote{https://www.euclid-ec.org} and  DESI\footnote{https://www.desi.lbl.gov} to all sky 21cm radio surveys like SKA\footnote{https://www.skatelescope.org}, Chime\footnote{https://chime-experiment.ca} and Tianlai\footnote{http://tianlai.bao.ac.cn}, where we will reach petabytes of data, it is clear that the future of cosmology will require fast and efficient algorithms for extracting scientifically meaningful information from the wealth of data collected.

A common statistical tool for compressing the spatial information of the galaxy distribution is the $N$-point correlation functions (or its Fourier counterpart the poly-spectra). 
Even at 3rd order statistics we have seen their usefulness in constraining the statistical bias between the distribution of galaxies and dark matter \citep[e.g., ][]{1993ApJ...413..447F,1998ApJ...503...37J,1999ApJ...521L..83F,2000MNRAS.313..725S,2001ApJ...546..652S,2002MNRAS.335..432V} as well as to place constraints on the amount of primordial non-Gaussianity in the cosmic microwave background \citep[CMB; ][]{2003ApJS..148..119K,2016A&A...594A..17P}.

The first configuration space galaxy 3-point correlation function (3pCF) measurements were made by \citet{1977ApJ...217..385G} and subsequently by \citet{1991ApJ...383...90G}, where they probed the {\em hierarchical clustering ansatz} \citep{1980lssu.book.....P} with $\mathcal{O}(1000)$  galaxies. However, more recently the 3pCF has been measured using hundreds of thousands of galaxies and used to place constrains on the Halo Occupation Distribution (HOD), by breaking degeneracies between galaxy bias and the amplitude of density fluctuations \citep{2015MNRAS.449L..95G}, it has shown promise at discriminating different gravity models \citep{2016A&A...592A..38S}, helping to constrain non-Gaussianity in the galaxy distribution \citep{2010JCAP...07..002N} and characterizing the turbulence of the interstellar medium \citep{2018ApJ...862..119P}.

Regarding large scale cosmology, \citet{2017MNRAS.468.1070S,2017MNRAS.469.1738S} made a detection of the Baryon Acoustic Oscillations (BAO) in the 3rd order spatial clustering statistics of the Sloan Digital Sky Survey (SDSS) galaxies. Although an earlier tentative detection of the 3pCF BAO was claimed by \citet{2009MNRAS.399..801G} who used the SDSS DR7 sample of Luminous Red Galaxies.

Naively, the CFs require computing the distance from every galaxy to every other galaxy in an $N_{gal}^2$ operation for the 2pCF or $N_{gal}^3$ for the 3pCF. However, there are some algorithmic methods that can speed up this kind of computation.

Arranging the data in a hierarchical structure known as a `tree', allows fast distance matching to be performed. Particularly k-d trees have been utilized for CFs in codes such as \texttt{Ntropy} \citep{2001misk.conf...71M,2007ASPC..376...69G} and \texttt{KSTAT} \citep{2018ascl.soft04026S, 2016A&A...592A..38S}.

Other novel methods have been developed to quickly measure the higher order statistics, including the
position-dependent power spectrum \citep{2014JCAP...05..048C} and multipole expansions \citep{2004ApJ...605L..89S}. More recently \citet{2015MNRAS.454.4142S,2016MNRAS.455L..31S} have developed a method based on Fourier transforms and spherical harmonics that can be combined to form the multipole coefficients of the 3pCF. 

In this work we will present a new algorithm for computing spatial correlations that is based on the concept of a {\em graph database}. A graph database  is a type of NoSQL database, {\em i.e.} it does not rely on the data being described in a tabular, relational format. Rather, a graph database, or more specifically a graph-oriented database uses graph theory to store, map and query relationships of the data.

The algorithm that we propose makes no approximation or data compression and is designed to measure all triplet configurations up to large cosmic scales beyond the BAO distance. 

The outline of this paper is as follows.
In \S\ref{sec:corr} we briefly review spatial correlation analysis. We detail the working of our new algorithm and explain the basic concepts behind graph databases in \S\ref{sec:algo}. In \S\ref{sec:test} we test our new algorithm, performing benchmark tests for speed and scalability. We also apply our algorithm to observational data in two example use cases; i) the large scale 3pCF ii) the four-point correlation function (4pCF). We then conclude in \S\ref{sec:conc}.

\section{Correlation Functions}
\label{sec:corr}
The spatial distribution of galaxies encodes a wealth of cosmological information. In an effort to condense the information of millions of 3D galaxy positions into a manageable form we rely on correlation functions. 

The two-point correlation function (2pCF) is defined as
\begin{equation}
\xi(\vec{r})=\left<\delta(\vec{x})\delta(\vec{x}+\vec{r})\right>_x,
\end{equation}
where $\delta$ is the density contrast, related to the density as $\delta\equiv\rho/\bar{\rho} - 1$ and $\left<..\right>$ denotes spatial averaging over $\vec{x}$. 

In practice the 2pCF is calculated using estimators, the most popular of which being the ``Landy-Szalay'' estimator \citep{1993ApJ...412...64L};
\begin{equation}
\xi(\vec{r})=\frac{DD(\vec{r})-2DR(\vec{r})+RR(\vec{r})}{RR(\vec{r})},
\label{eq:2pcf}
\end{equation}
where $DD$ is the number of data--data pairs, $DR$ the number of data-random pairs, and $RR$ is the number of random--random pairs, all  separated by a displacement vector $\vec{r}$ and properly normalised. The number of random particles used to define the unclustered reference sample is typically 20 times the number of data particles. This is done to reduce statistical fluctuation due to Poisson noise in the random pair counting. In the anisotropic analysis we decompose the vector $\vec{r}$ into a length component $s$ and the angle $\theta$ between the pair vector and the line-of-sight direction. 

\subsection{Three-point correlation function}
The 3pCF is defined as the joint probability of there being a galaxy in each of the volume elements $dV_{1}$, $dV_{2}$, and $dV_{3}$ given that these elements are arranged in a configuration defined by the sides of the triangle, ${\bf r}_{1}$, ${\bf r}_{2}$,  and ${\bf r}_{3}$. The joint probability can be written as
\begin{equation}
\begin{split}
    dP_{1,2,3}=\bar{n}^{3}[1&+\xi({\bf r}_{1})+\xi({\bf r}_{2})+\xi({\bf r}_{3})+\\
&+\zeta({\bf r}_{1},{\bf r}_{2},{\bf r}_{3})]dV_{1}dV_{2}dV_{3}.
\label{eq:3PCF}
\end{split}
\end{equation}
The expression above consists of several parts:  the sum of 2pCFs for each side of the triangle, $\zeta$, the full three-point correlation function, and $\bar{n}$  the mean density of data points.  We utilise the 3pCF estimator of \citet{1998ApJ...494L..41S}, 
\begin{equation}
\zeta = \frac{DDD - 3DDR + 3DRR - RRR}{RRR},
\label{eq:salay}
\end{equation}
where each term represents the normalised triplet counts in the 
data (D) and random (R) fields that satisfy a particular 
triangular configuration of our choice.
\begin{figure}
\includegraphics[width=\columnwidth]{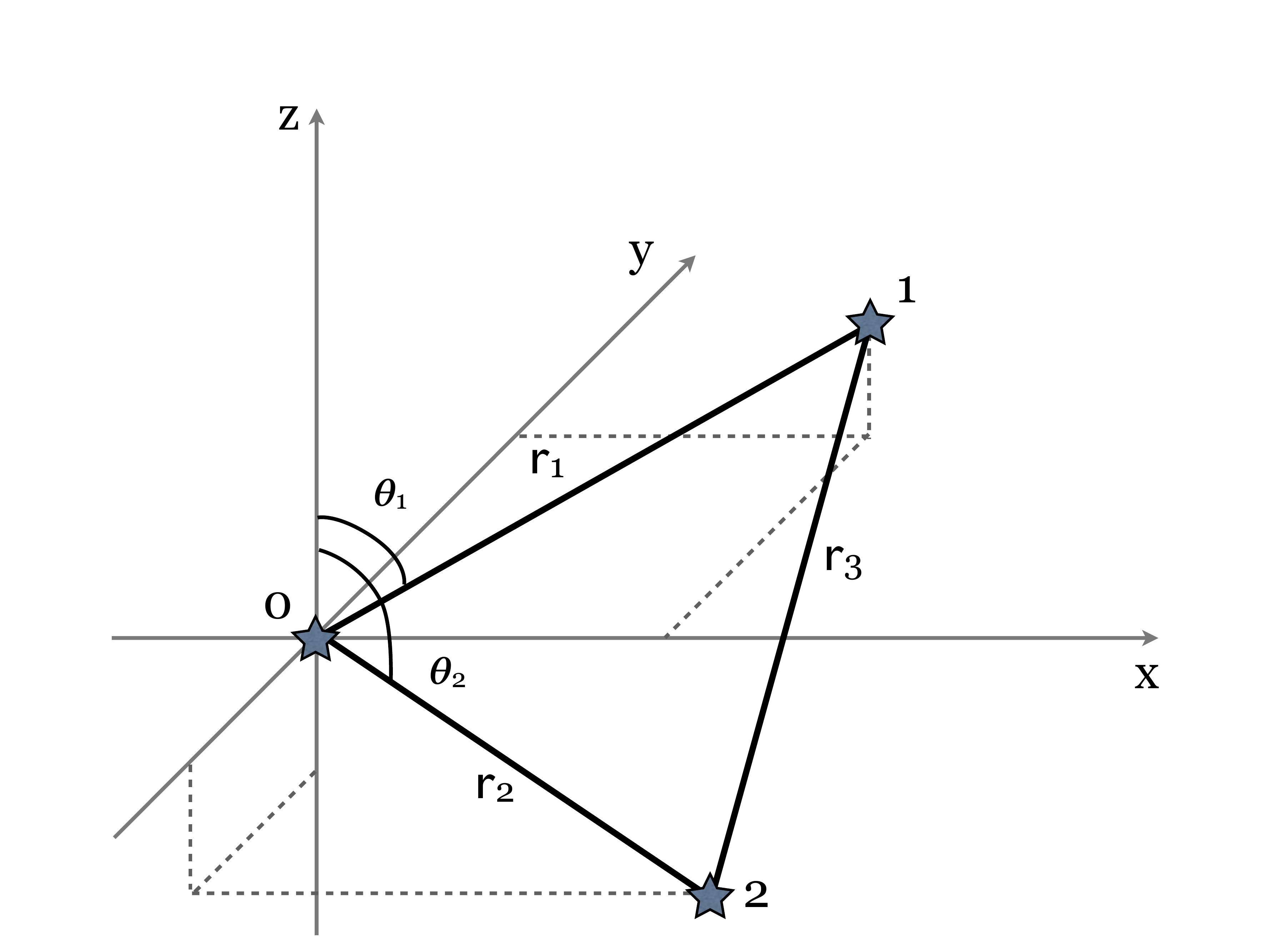}
\caption{This illustration shows the parameterization of a data triplet in anisotropic coordinates. Triangles are defined by 5 parameters: $(r_1,r_2,r_3,\theta_1,\theta_2)$.}
\label{fig:triangle}
\end{figure}

The 3pCF is a function of the three sides of the triangle $(r_1,r_2,r_3)$ and additionally it may also be computed in the anisotropic case, thus introducing two angles relative to the line of sight, $\theta_1$ and $\theta_2$. The triangular configuration parameters can be see in Figure \ref{fig:triangle}. 

We could imagine, for example, performing an analysis in the anisotropic 5-parameter space ($r_1,r_2,r_3,\theta_1,\theta_2$) with corresponding (20, 20, 20, 10, 10) equally spaced bins. Considering only legal triangles and symmetries, the number of possible bins is $\approx$200,000. The large number of possible bins makes covariance estimation a serious issue for higher order statistics. Thankfully recent work has shown the possibility to reduce the number of bins using various compression schemes
\citep{2018arXiv180602853G,2018MNRAS.476.4045G,2018arXiv180611147C,2018arXiv181112396C}.

\section{Algorithm Design}
\label{sec:algo}
\subsection{Graph Database}

A graph database does not rely on the data being described in a tabular or relational format, as with more traditional database structures. Rather, a graph database uses graph theory to store, map and query relationships of the data.

\begin{figure*}
	\includegraphics[trim={4cm 5cm 4cm 2cm},clip,width=1.1\columnwidth]{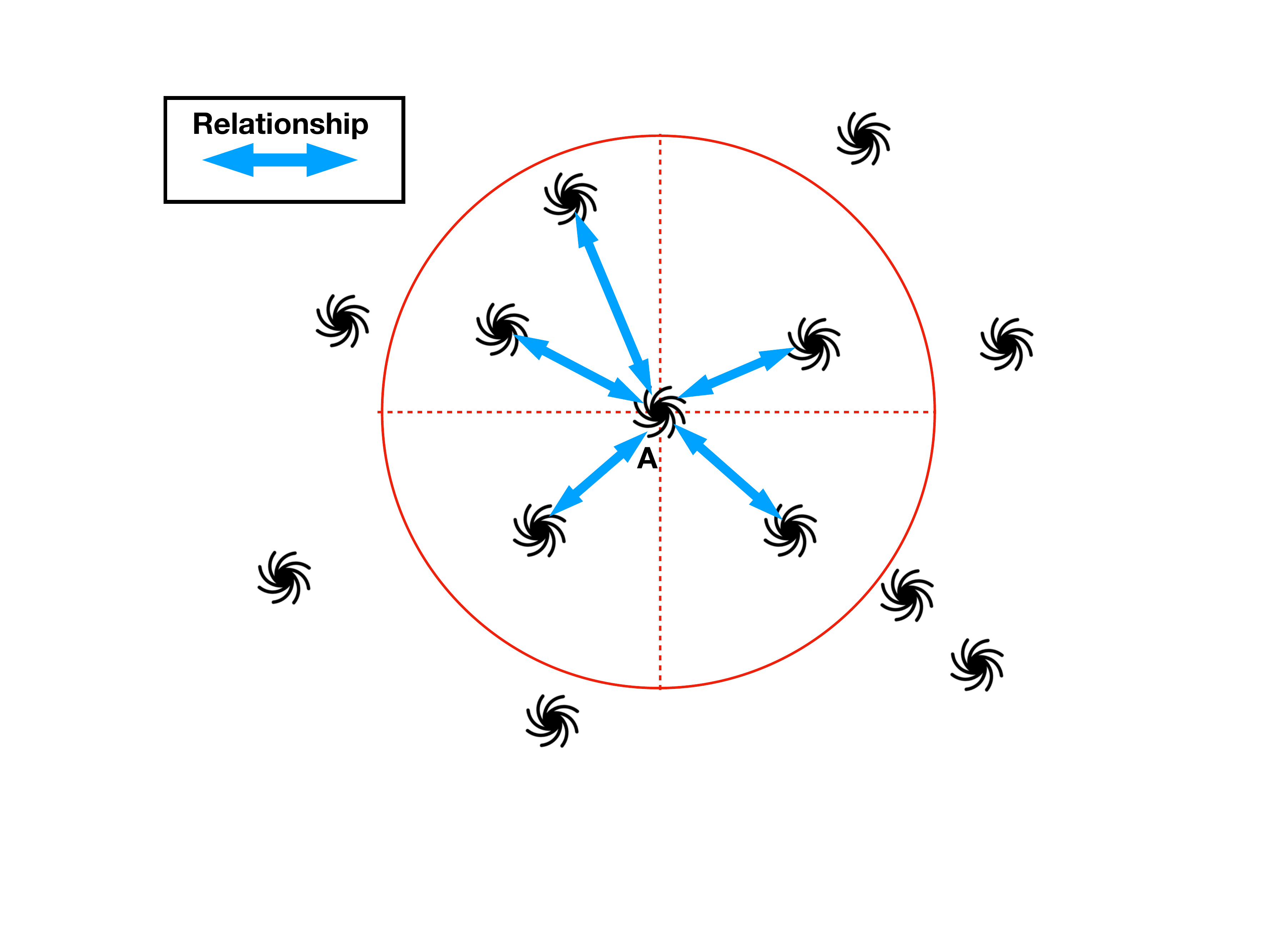}
    \includegraphics[trim={5cm 3cm 4cm 5cm},clip,width=1.1\columnwidth]{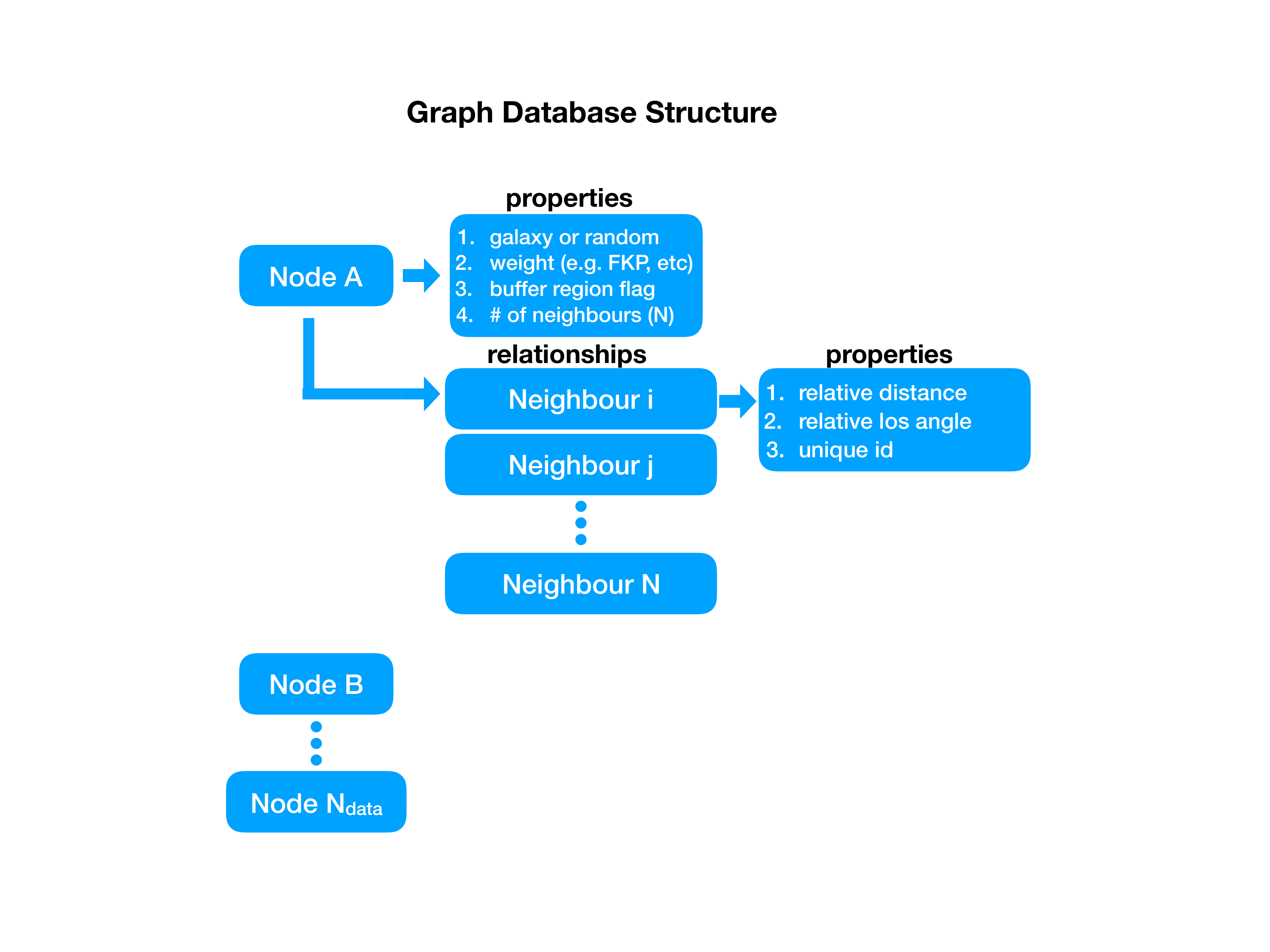}
    \caption{{\em Left:} This is our definition of a node. A node is a data point, (e.g. galaxy or random) that contains a list of relationships between itself and its nearest neighbors within a predefined length scale, $r_{max}$. In this example Node A has 5 relationships to other data points, but has no relationship to data points beyond $r_{max}$. {\em Right:} This is the structure of the graph database. Each data point is a Node containing certain properties about itself. It also contains a list of relationships. In our case this is a list of neighbor points within a distance of $r_{max}$ from Node A. Each neighbor also has properties, these include the distance information to Node A and optionally the angular information relative to the line of sight. The also contain the unique ID of the data point.}
    \label{fig:node}
\end{figure*}

Each data point is called a node and has a number of associated properties. 
On the right side of Figure \ref{fig:node} we see the main elements of the graph. In our work, the important properties of a node are 1) if the point is a galaxy or random 2) any weight associated with the data point {\em e.g.} FKP weights  \citep{1994ApJ...426...23F}, angular systematic weights \citep{2012MNRAS.424..564R}, etc 3) the number of neighbour points within a fixed radius 4) if the data point is within a buffer region. The buffer region (4) is only required if the data has been decomposed into multiple domains which will be discussed later. The number of neighbours (3) is required solely to facilitate dynamical memory allocation.

The graph is constructed by visiting each data point and building a list of relationships to its neighbors within a distance, $r_{max}$, see left panel of Figure \ref{fig:node}. The list of neighbors can be obtained quickly using a kd-tree\footnote{We have used the open source solution called \texttt{kdtree2} from \citet{2004physics...8067K} \url{https://github.com/jmhodges/kdtree2}}.

As we can see in the right panel of Figure \ref{fig:node}, each node relationship contains the distance information and optionally the angle relative the line of sight direction, if anisotropic correlation analysis is required. It also contains the unique ID of the data point. In an effort to be memory efficient, we bin the distance and angular information immediately thus turning a single/double precision floating point number into a integer. Since the number of bins is typically of order 100, we can save it as a 8byte integer ($2^8=256$ unique values), reducing the required memory significantly.

\subsection{Querying the database: specific configurations}
\label{sec:query}

Unlike for more traditional databases, graph database queries work on relationships and properties thereof. As a simple example if we want to compute the 2-point correlation function we can see that it can be obtained immediately by counting all possible relationships where the distance property of the relationship matches our desired scale. This list of relationships can be queried further to find which ones correspond to the pairs of data (DD), the pairs of random points (RR) and the mixture of both (DR). 

Specific $n$-tuples can be computed by querying the relationships and relationships of relationships such that the distances satisfy $r_1,r_2,...,r_n$ and that the initial and final data point have the same unique ID, thus {\em closing} the tuple.

\subsection{Querying the database: all configurations}
In some circumstances we may wish to compute all binned configurations of an $n$-tuple correlation function. 
As an example in the 3pCF, we would have to count all possible triplets of points with the sole criteria that $r_1,r_2,r_3<r_{max}$. Thankfully this calculation can be performed rather efficiently by making use of a nice geometrical property of our database. 

\begin{figure*}
	\includegraphics[trim={7cm 3cm 7cm 3cm},clip,width=0.8\columnwidth]{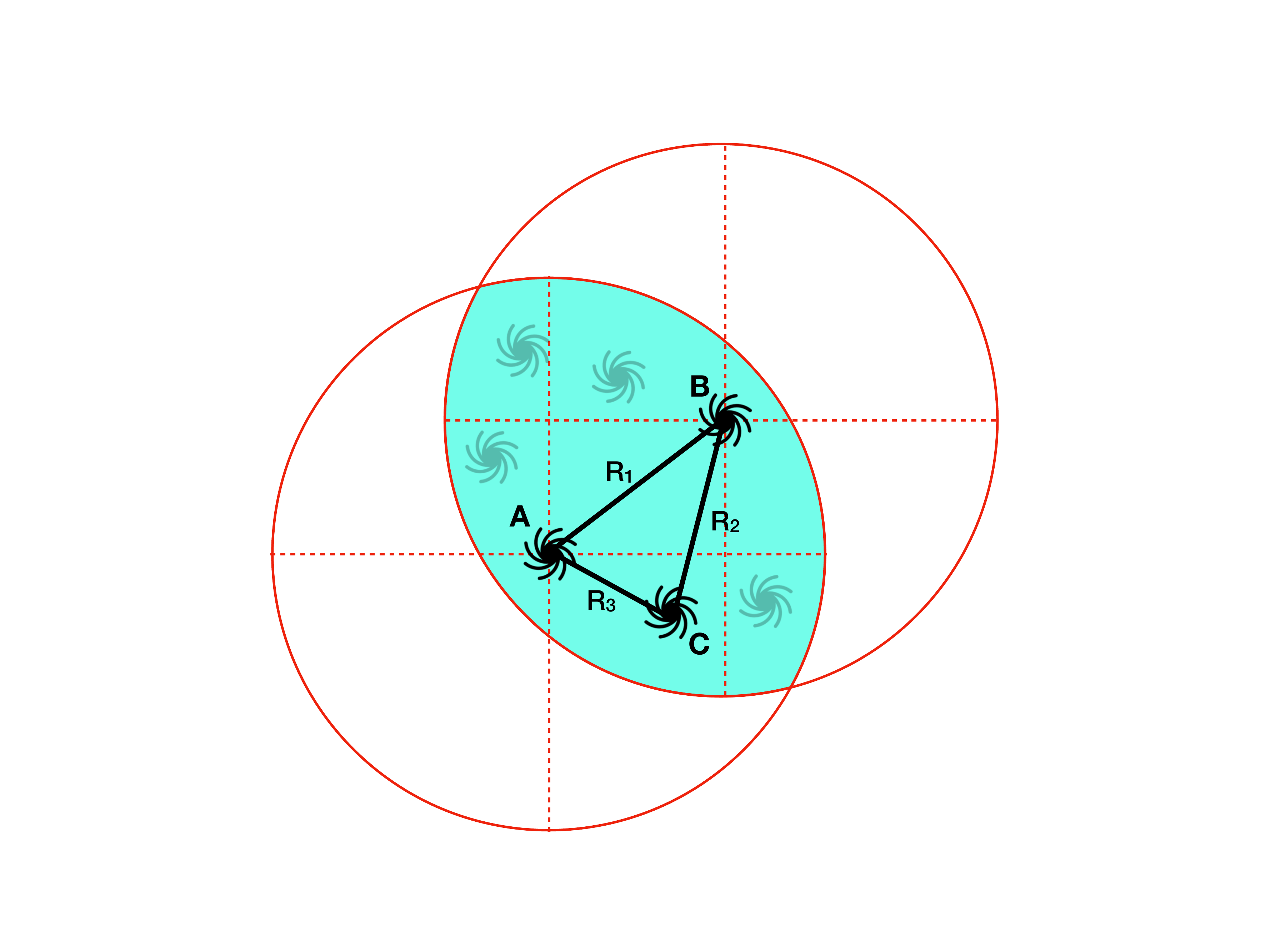}
	\includegraphics[trim={1cm 2cm 0cm 4cm},clip,width=1.3\columnwidth]{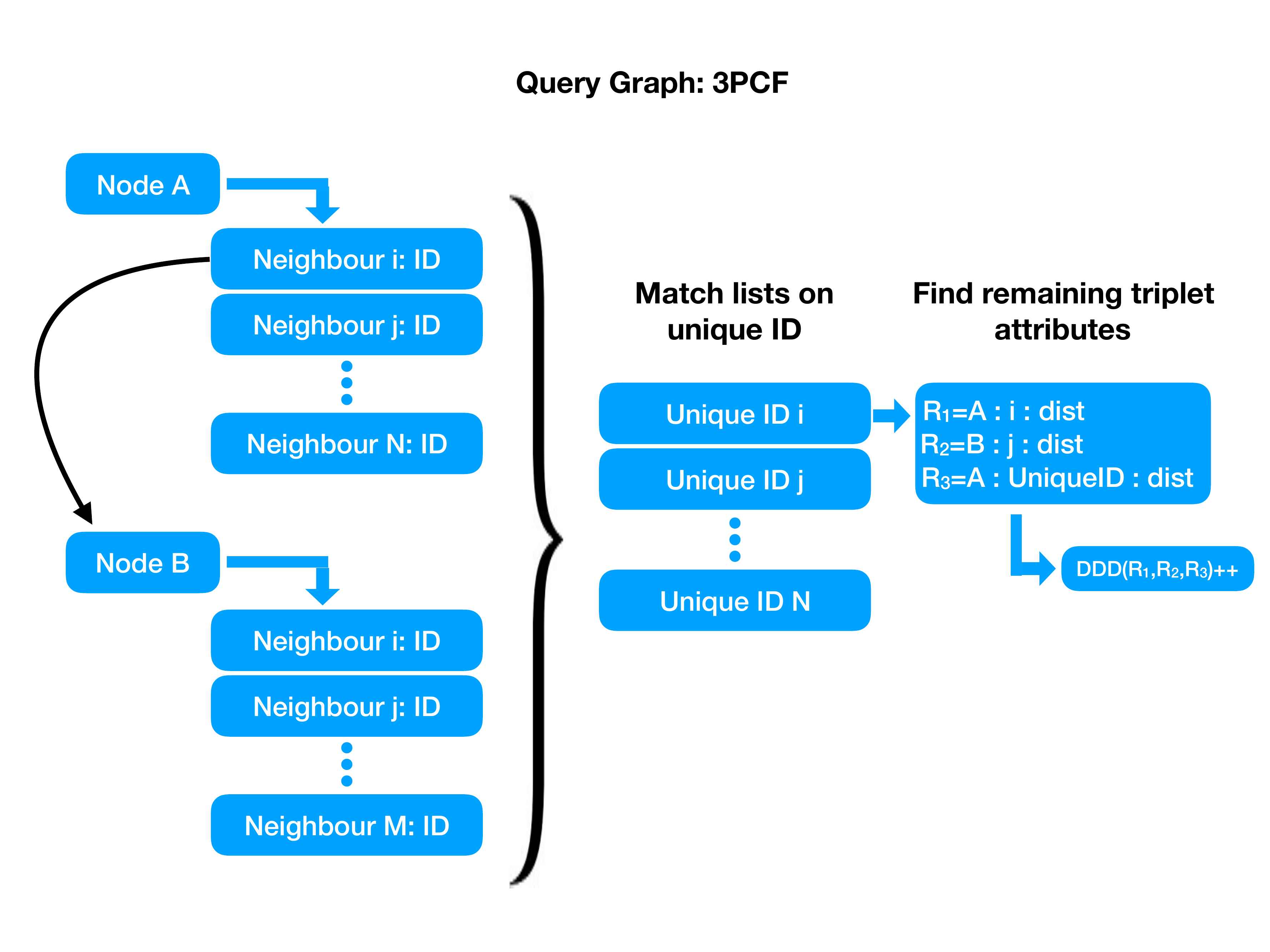}
    \caption{{\em Left:} The intersection (shaded region) between two spatially overlapping nodes, A and B, contains a list of points satisfying our triangular configuration constraints, e.g. node C.  {\em Right:} This shows the computational structure of the graph query. Node A is opened and then recursively its relations are opened. The first neighboring point is identified as Node B which is then opened. These two open Nodes contain a list of neighbor points, which are then matched according to their unique ID. This resulting intersection contains all possible triplets satisfying the condition $r_1,r_2,r_3<r_{max}$. }
    \label{fig:graph}
\end{figure*}

Between a node and one of its relationship data points there can be defined a region containing all possible triplets associated to the original pair with configuration $r_1,r_2,r_3<r_{max}$. This is a simple geometrical property and can be more easily visualized in the left panel of Figure \ref{fig:graph}.

In the right panel of Figure \ref{fig:graph} we can see illustratively how the triplets are counted. We firstly visit each Node, eg Node A and open its list of relationships. We then open the Node corresponding to the first relationship (neighboring point). This Node (e.g. Node B) also contains a list of relationships. If we match each of these lists on the unique ID of each data point, then we will have a new list containing all possible triplets, where two corners are Nodes A and B and the side lengths satisfy the constraint $r_1,r_2,r_3<r_{max}$. The union of two ordered sets can be performed very quickly, thus we initially sort the neighbor lists by their unique ID. We can see why this is computationally efficient, because the length of the 3rd side of the triangle does not need to be calculated, since all distance between data pair have been already been measured. This is the main factor in this algorithm, {\em all distances are precomputed and saved into memory}.

Due to the precomputation of the distances and the specific design of the graph database, it becomes trivial to go beyond the 2- and 3pCF to arbitrary order in the $N$-point statistics. This may allow us to estimate directly the covariance matrices of lower order statistics, since for example the cov$(\xi)$

\subsection{Domain Decomposition}
\label{sec:domain}
\begin{figure}
    \centering
	\includegraphics[trim={2cm 1cm 2cm 1cm},clip,width=0.9\columnwidth]{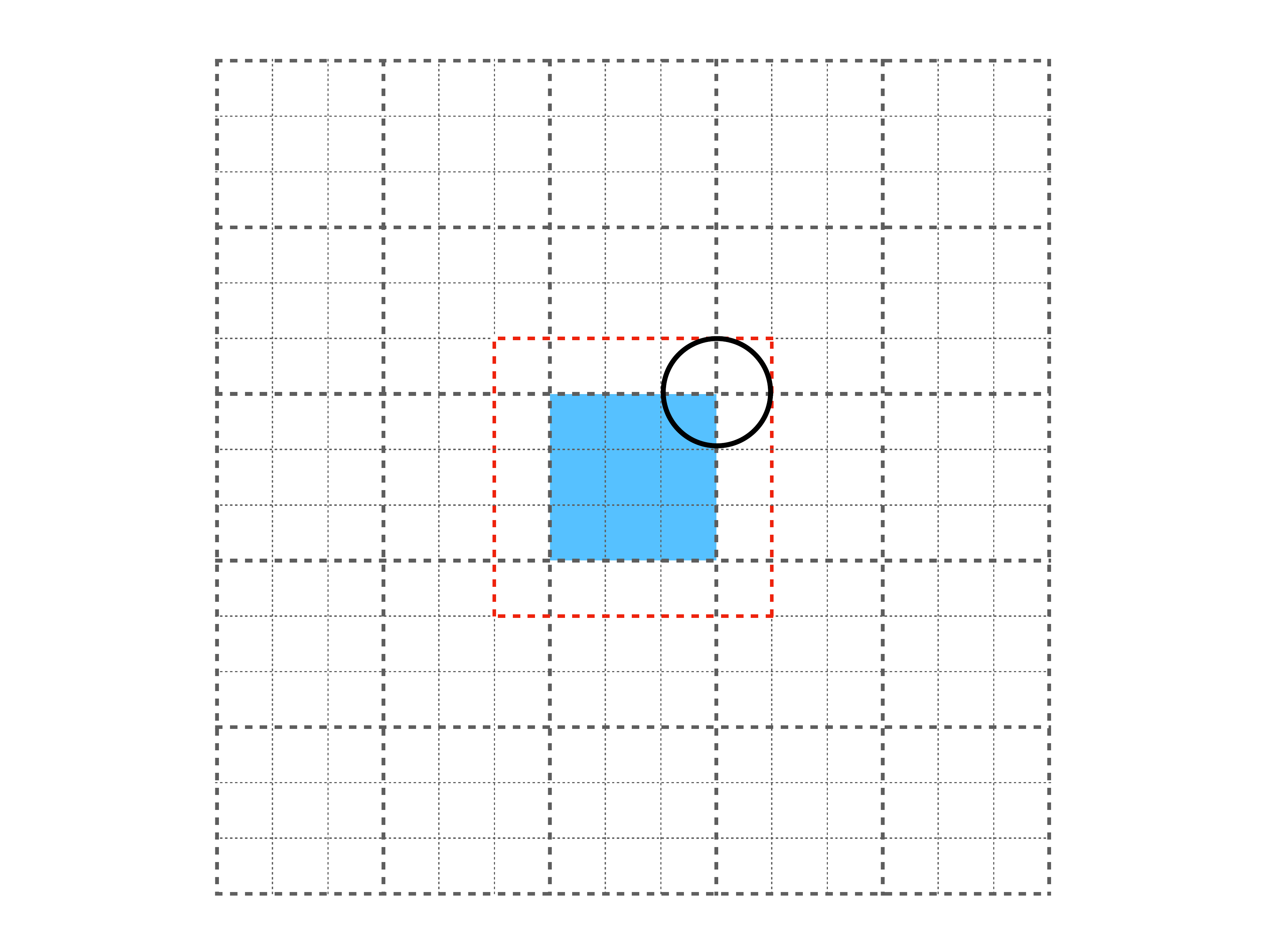}
    \caption{This is the domain decomposition scheme for our algorithm. In this 2D example the space is initially split into 25 areas containing equal number of data points. In accounting for the range of the correlation analysis we add a buffer region (red dashed line) around the initial domain (blue region). The black circle denotes the maximum scale of our analysis, $r_{max}$.}
    \label{fig:domain}
\end{figure}

The graph database can become memory intensive, e.g. for 10 million points with a number density, $\rho\approx10^{-4}$ Mpc$^3h^{-3}$ and analysis scale $r_{max}=150$ Mpc/$h$, this could reach almost 1TB of memory. Thus if we want to query such a database quickly it would ideally be loaded into RAM memory. We could imagine constructing a large database (e.g. SQL, etc) that can be externally called or large HDF5 files that can be optimally constructed for fast random access calls. However, we decided to opt for a distributed memory scheme that would allow different parts of the graph database to reside on many compute nodes separately. Since no compute node can hold all of the database, we use a domain decomposition to spatially partition the data nodes over $N$ compute nodes. This can be seen in Figure~\ref{fig:domain}. Each domain must also include a buffer region since nodes outside the domain may be required for the correlation function analysis. Therefore the buffer region extends a distance $r_{max}$ in each spatial direction. With this scheme we can be sure that all relevant data nodes are available in memory for the required domain correlation analysis. 

\subsection{Computational Structure}
\begin{figure}
	\includegraphics[trim={1cm 0cm 3cm 2cm},clip,width=1.05\columnwidth]{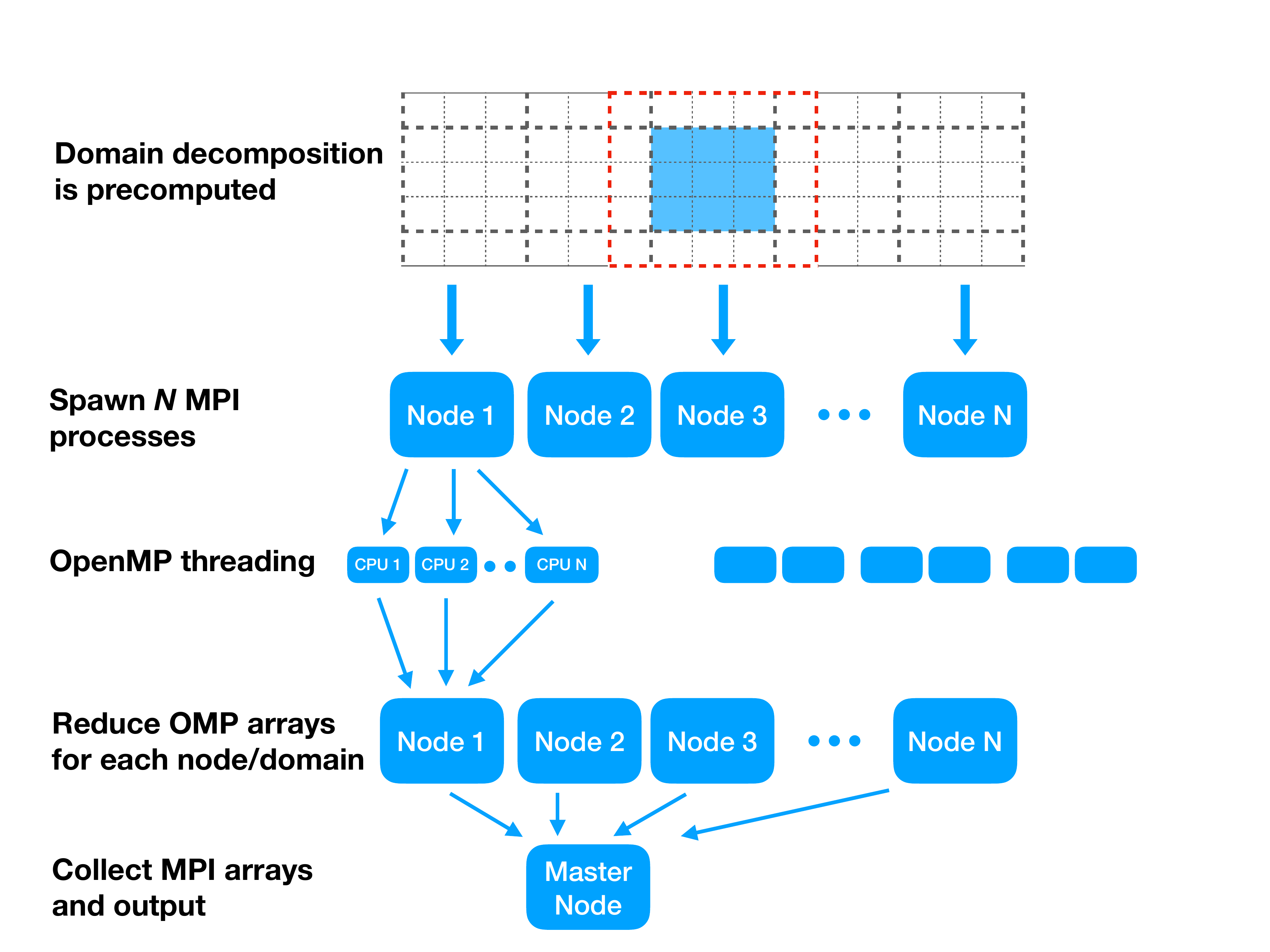}
    \caption{This is the overall computational structure of our algorithm (from top to bottom). The data is initially split into N files according to our domain decomposition scheme. Each domain is then loaded into memory on a separate compute nodes. Each node then open several OpenMP threads for distributing the task of graph database construction and then graph querying. The OpenMP threads are reduced, collecting the local pair and triplet counts, then the MPI processes are collected on the master Node for final calculation and result output.}
    \label{fig:comp_struct}
\end{figure}

In Figure \ref{fig:comp_struct} we show the computational structure of the algorithm. The $N$ precomputed domains are read by $N$ compute nodes, {\em i.e.} one domain per compute node. This can be achieved using MPI as shown in the figure, or since each domain is independent, the code could be run many times with 1 MPI thread and the user then gathers the results individual for each of the $N$ domains. For now we will consider the MPI case.

Each MPI task loads the particle data into memory and proceeds to construct a kd-tree. Once the kd-tree has been constructed we spawn a number of OpenMP threads for parallel computation. The kd-tree is then queried for every point, and their $N$ neighbors within a distance $r_{max}$ having their distance computed and, optionally, the angle between the connecting vector and the line-of-sight direction. The distance and angle can be saved for each neighbor of each data point as an 8byte integer in a structure, as shown in the right panel of Figure \ref{fig:node}. This is the construction of the graph database.  

Once the graph database has been constructed, the code proceeds to query the graph and measure all the triplet configurations, as discussed above. Again we invoke a number of OpenMP threads to speed up the calculation on each compute node. 

The MPI tasks are then reduced to one master node where the $DDD$, $DDR$, $DRR$, $RRR$ counts are each combined together and normalized by the total possible number of triplet counts (including weights). 

In this work we present results from our own custom graph database written in modern \texttt{FORTRAN} and implementing both  \texttt{OpenMP} and \texttt{MPI} protocols. However we have also tested the popular open source graph database \texttt{Neo4j}\footnote{https://neo4j.com} finding similar performance.

\section{Testing and Benchmarking}
\label{sec:test}

\begin{figure*}
\includegraphics[width=\columnwidth]{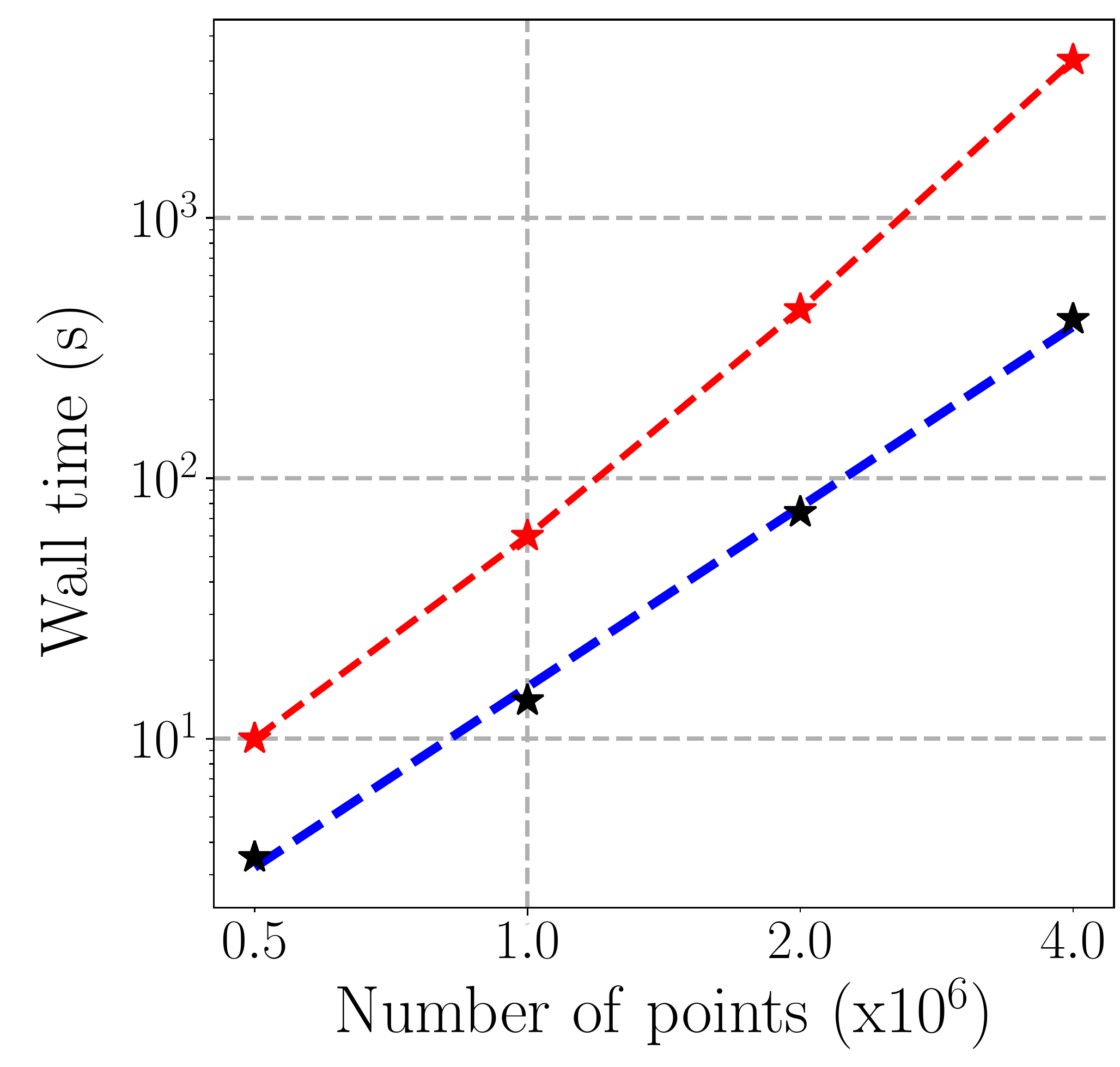}
\includegraphics[width=\columnwidth]{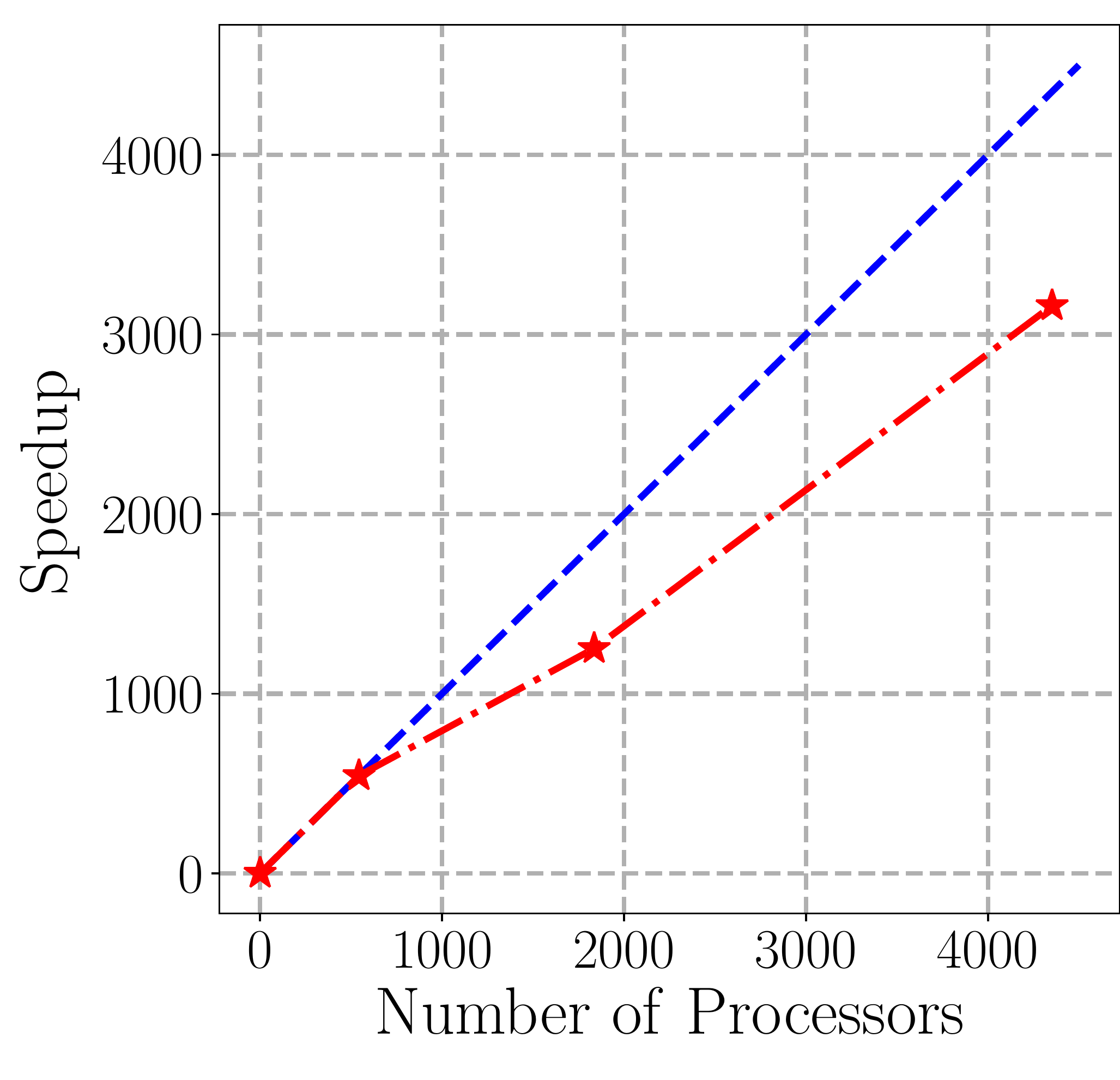}
\caption{\label{fig:test} {\em Left:} The computational wall clock time for a single call of the 3pCF is shown as a function of number of data points. The red points are computed using 4 MPI threads with \texttt{KSTAT} while the black points are from the Graph database code. The blue dashed line  corresponds to an $N^{2.3}$ relation. {\em Right:} The computational speedup scaling with the number of processors. An ideal case with perfect load-balancing would expect to give 1 to 1 scaling (blue dashed). The graph database produces reasonable scaling up to several thousands of processors (red stars).}
\end{figure*}

For the purposes of benchmarking we focus on small scales and count all possible triangular configurations within $r_1,r_2,r_3<30$ Mpc. We use 4 random data samples with a varying number of data points, $N_{data}$, and densities while keeping the volume constant at 1 Gpc$^3$. We use samples with $N_{data}$ = $\{$0.5, 1.0, 2.0, 4.0$\}$ million points. In Figure \ref{fig:test} we compare our new algorithm (black points) with the kd-tree algorithm, \texttt{KSTAT} (red points). The new algorithm is significantly faster than \texttt{KSTAT} and the scaling also shows a gentler slope with the number data points with an approximate scaling relation of $N^{2.3}$. 

The reason for the discrepancy between the two algorithms is due to the efficient information handling of the graph database. In the case of \texttt{KSTAT}, the algorithm must compute the anisotropic angles for every pair and triplet of data. However, the graph database only requires this calculation to be performed on pairs of data points and then to compute 3-point statistics it simply retrieves the relevant information from the database of relationships as described in \S\ref{sec:query}.

\subsection{Scalability}
We investigate the performance of the algorithm with increasing parallelisation. Adopting a hybrid MPI + OpenMP scheme we perform a test using multiple Intel Xeon Phi 7250 nodes, each comprising 68 computational cores and 96 GB of RAM. Thus it is natural to set the number of OpenMP threads as 68 and create (and query) the graph database for each domain on separate compute nodes, as illustrated in Figure \ref{fig:comp_struct}.


In the right panel of Figure \ref{fig:test} we show how the wall clock time scales with number of processors. We find reasonable scaling up to several thousand processors. The departure from the 1 to 1 scaling is due to imperfect load-balancing {\em i.e.} the distribution of work over the individual processors. 

In galaxy clustering it is impossible to know {\em a priori} how much computational time will be required for different parts of the data, because of the complex survey geometry and the spatial clustering of the data itself. Thus if we want to avoid significant interprocessor communication we reply on an an initial domain decomposition scheme, which was described earlier in \S\ref{sec:domain}.

\subsection{Example case I: 3-point BAO}
We now investigate the performance of the algorithm on large scales. One of the goals of modern cosmology is to determine the expansion history of the Universe through distance measurements, which can be obtained by identifying the BAO `bump' in the 2pCF \citep{2005ApJ...633..560E,2012MNRAS.427.3435A}. It is also expected that this feature should be present in the higher order statistics. 

We consider Data Release 12 of the Sloan Digital Sky Survey's \citep[SDSS;][]{2011AJ....142...72E}  Baryon Oscillation Spectrioscopic Survey {\em Constant Stellar Mass} \citep[CMASS;][]{2012AJ....144..144B,2013AJ....145...10D,2015ApJS..219...12A} sample\footnote{https://data.sdss.org/sas/dr12/boss/lss/}. Specifically we use their publicly released observational catalogues for the Northern Galactic Patch ({\em i.e.} \texttt{galaxy\_DR12v5\_CMASS\_North}) and 150 {\em Quick-Particle-Mesh}  \citep[QPM; ][]{2014MNRAS.437.2594W} mock galaxy catalogues that mimic the observational geometry and selection effects. In the observational sample and in each mock catalogue there are approximately 500,000 galaxies and 2 million random points over the redshift range $0.43<z<0.7$. In transforming from redshift space to comoving Cartesian coordinates we adopt a flat  $\Lambda$CDM cosmology model with $h=0.7$ and $\Omega_m=0.3$.

\begin{figure*}
\includegraphics[trim={0cm 0cm 0cm 0cm},clip,scale=0.7]{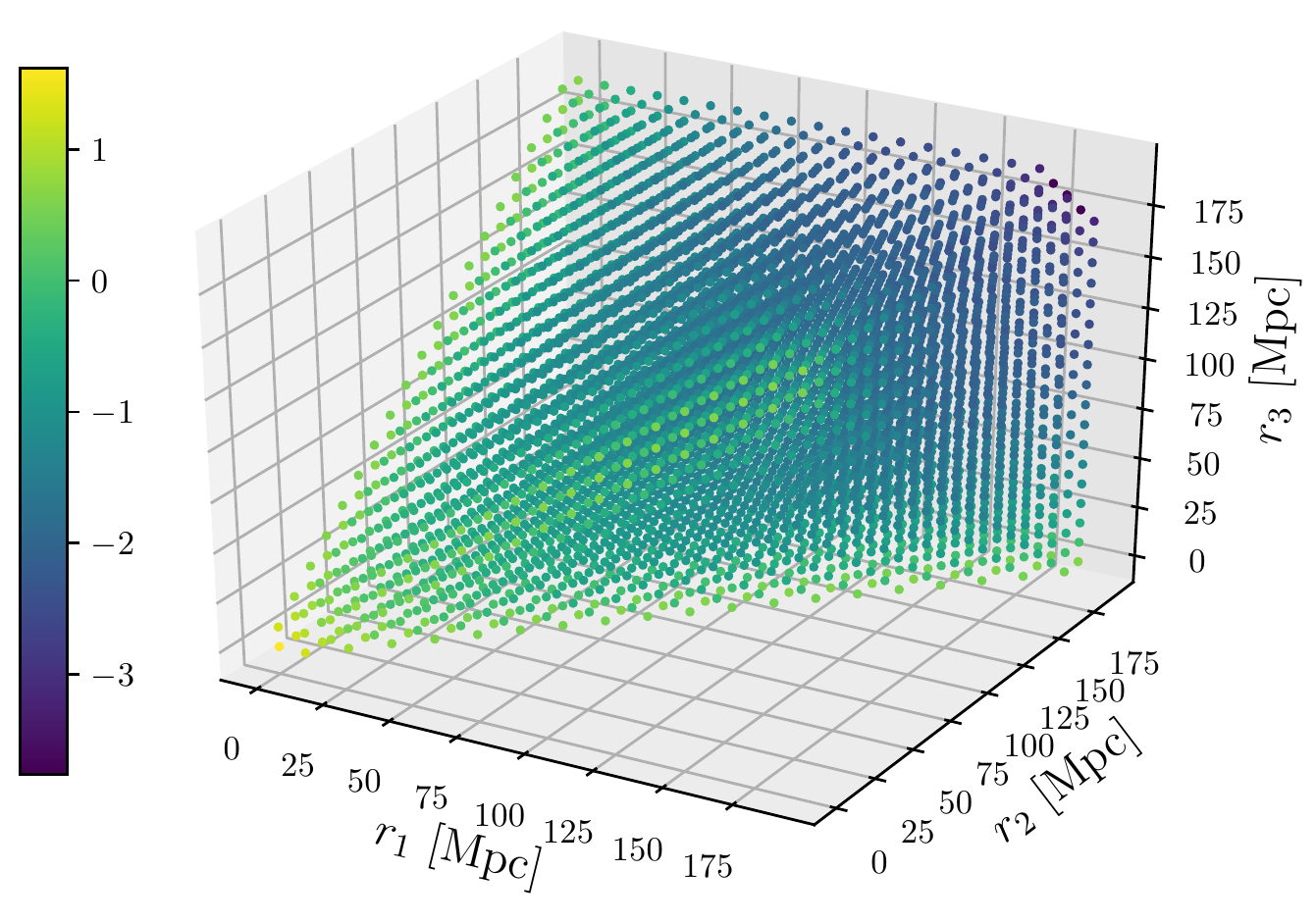}
\includegraphics[scale=0.38]{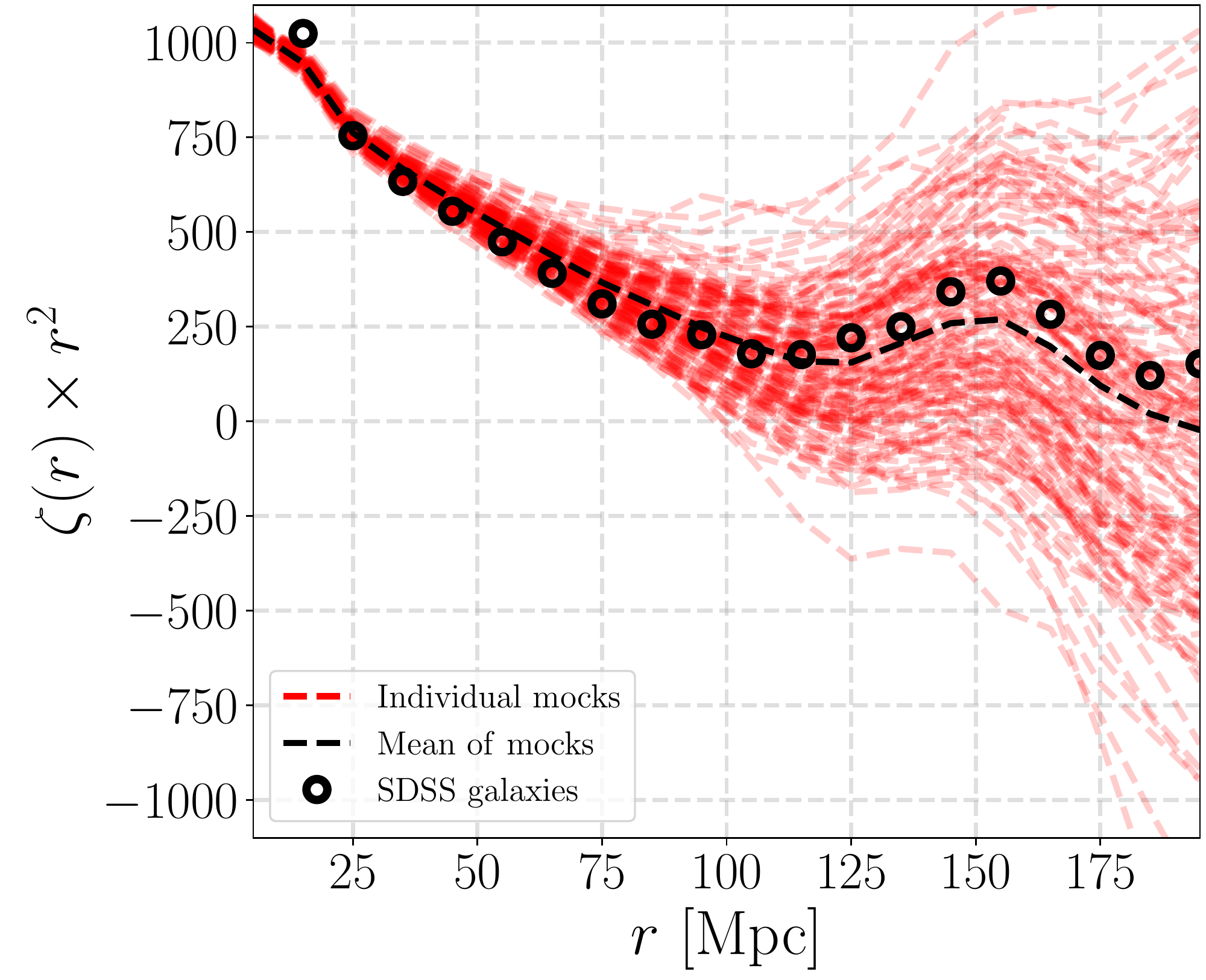}
\caption{\label{fig:BAO} The large scale 3pCF is shown for 150 mock SDSS DR12 CMASS samples. {\em Left:} The full space of $r_1,r_2,r_3<200$ Mpc configurations with $\Delta r=10$ Mpc binning. The colored points denoted the mean 3pCF value of the mocks for specific binned triplet configurations. {\em Right:} The equilateral 3pCF for for each mock catalogue (red) and for the mean of the 150 mock catalogues (black). Although there is significant scatter among the mock samples the mean is very smooth and exhibits a clear BAO `bump' at $\sim150$ Mpc. }
\end{figure*}

We proceed to measure the 3pCF beyond the BAO scale out to 200 Mpc. In the left panel of Figure \ref{fig:BAO} we show the 3pCF for the mean of the mock catalogues and for all possible binned triangular configurations with $r_1,r_2,r_3<200$ Mpc. This information is difficult to display in 2-dimensions so for clarity we can look at slices or projections through this space. In the right panel of Figure \ref{fig:BAO} we display the equilateral configurations for each mock individually (red lines), for the mean of mock catalogues (black dashed line) and for the observational sample (back circles).  Although there is significant scatter among the mocks, the BAO `bump' is clearly visible in the mean of the mock samples and also in the observational sample. 

Each galaxy catalogue $DDD$ count took just 515 seconds on a 68 core Intel Xeon Phi processor. Although the full calculation of the $RRR$ counts took considerably longer at $\sim$4 hours on 27 Intel Xeon Phi nodes (1,836 computational cores).

\subsection{Example case II: 4-point function}

Following hot on the heels of \citet{1978ApJ...221...19F}, we compute the 4-point correlation function from the mocks and observational catalogues presented in the previous section. 

For simplicity we adopt the following estimator for the 4pCF, 
\begin{equation}
    \eta({r_1,r_2,r_3,r_4})=\frac{DDDD}{RRRR}-1
\end{equation}
where DDDD and RRRR represent quadruplets of data points that satisfy the criteria $r_1,r_2,r_3,r_4$ for each side of the quadrilateral. Although in general the sides may be vectors and the 4 points may not occupy a single plane in 3D in which case the most general quadruplets have {\em skewed quadrilateral} configurations. 

\begin{figure}
    \centering
    \includegraphics[width=\columnwidth]{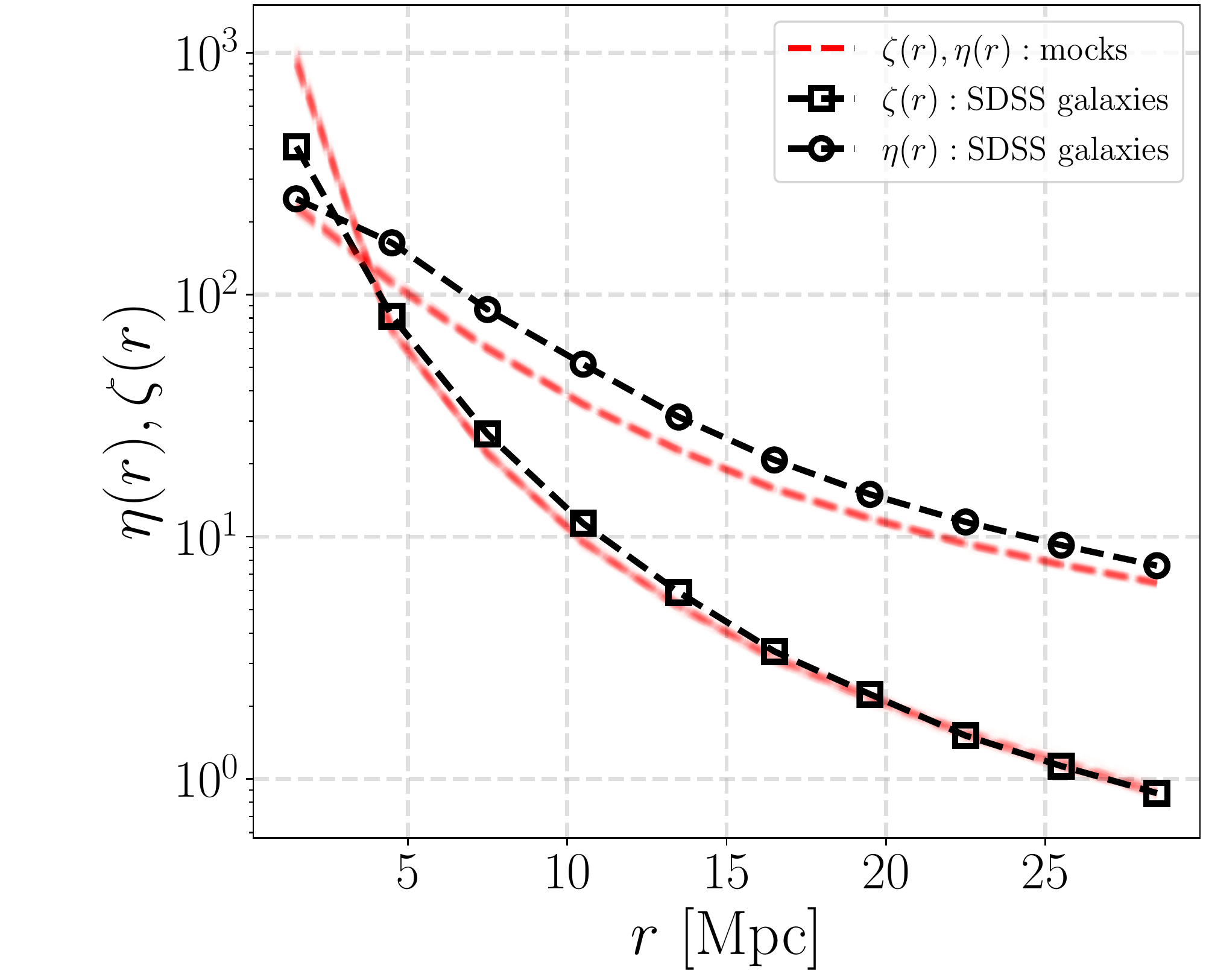}
    \caption{\label{fig:4pt}The equilateral 3pCF ($\zeta$) and the box 4pCF ($\eta$) are shown on scales $0<r<30$Mpc. Measurements are shown for 150 individual mock galaxy catalogues (red) and for the SDSS DR12 CMASS galaxies. The dispersion between the mocks is too small to visualize in this scale.}
\end{figure}

In this example we restrict ourselves to general equi-sided quadrilateral configurations and varying the side length from $0<r<30$ Mpc in 10 equally spaced bins. We proceed to measure the 4pCF in the observational catalogue and in each of the 150 mock galaxy catalogues. In Figure \ref{fig:4pt} we show the equilateral 3pCF and the box 4pCF up to 30 Mpc. It is clearly noticeable that the 4pCF of the mocks and the observational galaxies are in tension, while the 3pCF shows much better agreement. This is not too surprising given the approximate nature of the QPM method, which is already known to breakdown for 3rd order statistics on mildly non-linear scales \citep{2016MNRAS.456.4156K}.

The computational time required for one catalogue was approximately 1 minute on 4 OMP threads, which included 1.8s to construct the graph, 9s to query the required triplet counts and a further 40s to obtain the quadruplets for the 4pCF.

\section{Conclusions}
\label{sec:conc}
We present a fast algorithm for the computation of all possible triplet configurations with $r_1,r_2,r_3<r_{max}$ of a discrete point set. 

Through benchmarking we demonstrate that the algorithm scales well with increasing number of data points up to millions and can easily handle the current cosmological data sets.

We show reasonable parallel scalability through initial domain decomposition and load-balancing. Although there may be room for improvement with a dynamical load-balancing scheme.

The BAO at 3rd order is presented showing the visual BAO peak structure in both mocks and observational data. As it was not the primary aim of this work, we save interpretation of the BAO signal for future work.

We also show for the first time the 4pCF of SDSS galaxies. Again, as it was not our primary goal to make cosmological inferences, we will leave the interpretation of the 4pCF to forthcoming work.

We optimized the code to run on the Cray CS500 system Nurion at the  Korea Institute of Science and Technology Information (KISTI) which comprises 570,020 compute cores. Running the code on a single catalogue of the SDSS BOSS DR12 sample containing $\sim$500,000 galaxies and 2 million random points we computed the full anisotropic 3pCF up to 200 Mpc. This calculation took $\sim$4 hours on 27 Intel Xeon Phi 7250 nodes (1,836 computational cores).

Finally, we present the publicly available code  
\texttt{GRAMSCI} (GRAph Made Statistics for Cosmological Information; \url{https://bitbucket.org/csabiu/gramsci}), under a GNU General Public License. 

\section*{Acknowledgements}
We thank Mijin Yoon for useful discussion and comments on the manuscript. 

We use the Nurion supercomputing cluster at the Korea Institute of Science and Technology Information (KISTI).

CGS acknowledges financial support from the National Research Foundation (NRF; \#2017R1D1A1B03034900)




\end{document}